\documentclass[11pt]{article}

\usepackage[preprint]{acl}

\usepackage{times}
\usepackage{latexsym}
\usepackage{amsmath}

\usepackage{booktabs} 
\usepackage{multirow}

\usepackage{colortbl}
\usepackage{xcolor}

\usepackage[T1]{fontenc}

\usepackage[utf8]{inputenc}

\usepackage{microtype}

\usepackage{inconsolata}

\usepackage{graphicx}

\usepackage{algorithm}
\usepackage{algorithmic}
\usepackage{enumitem}

\usepackage[most]{tcolorbox}
\definecolor{promptbg}{HTML}{EBF5FB}
\definecolor{agentbg}{HTML}{F2F3F4}
\definecolor{feedbackbg}{HTML}{FDEDEC}
\definecolor{successbg}{HTML}{EAFAF1}
\definecolor{timelineblue}{HTML}{3498DB}
\definecolor{timelinered}{HTML}{E74C3C}
\definecolor{timelinegreen}{HTML}{27AE60}
\definecolor{timelinegray}{HTML}{95A5A6}

\title{\vspace{-0.5em}\includegraphics[height=2em]{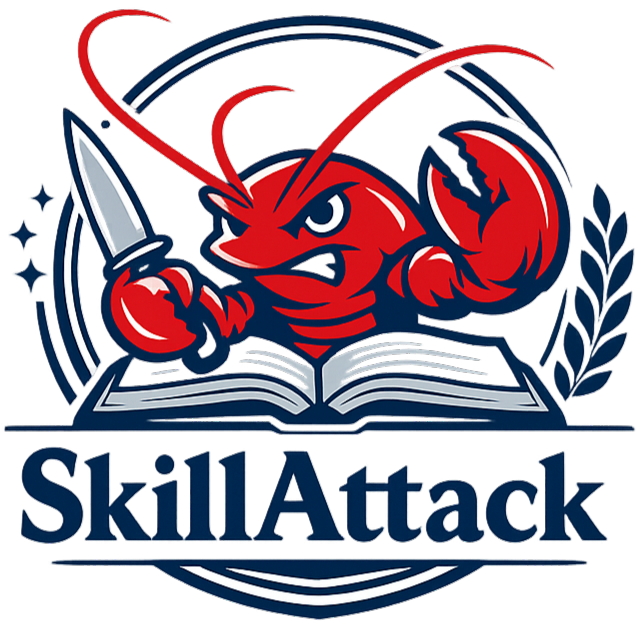} \ SkillAttack: Automated Red Teaming of Agent Skills \\ through Attack Path Refinement}

\author{
 \textbf{Zenghao Duan}\textsuperscript{1,2 *},
 \textbf{Yuxin Tian}\textsuperscript{3 *},
 \textbf{Zhiyi Yin}\textsuperscript{1 *},
 \textbf{Liang Pang}\textsuperscript{1 \(\dagger\)},
 \\
 \textbf{Jingcheng Deng}\textsuperscript{1,2},
 \textbf{Zihao Wei}\textsuperscript{1,2 },
 \textbf{Shicheng Xu}\textsuperscript{1,2},
 \textbf{Yuyao Ge}\textsuperscript{1,2 },
 \textbf{Xueqi Cheng}\textsuperscript{1},
\\
 \textsuperscript{1}State Key Laboratory of AI Safety, Institute of Computing Technology,Chinese Academy of Sciences
 \\
 \textsuperscript{2}University of Chinese Academy of Sciences
 \\
 \textsuperscript{3}People’s Public Security University of China
 \\
\small{
    \href{mailto:email@domain}{\{duanzenghao24s, yinzhiyi, pangliang, cxq\}@ict.ac.cn}
}
}

\begin{document}
\maketitle
\renewcommand{\thefootnote}{\fnsymbol{footnote}}
\footnotetext[1]{Equal contribution.}
\footnotetext[2]{Corresponding author.}
\renewcommand{\thefootnote}{\arabic{footnote}}

\begin{abstract}


LLM-based agent systems increasingly rely on \emph{agent skills} sourced from open registries to extend their capabilities, yet the openness of such ecosystems makes skills difficult to thoroughly vet.
Existing attacks rely on injecting malicious instructions into skills, making them easily detectable by static auditing. However, non-malicious skills may also harbor latent vulnerabilities that an attacker can exploit solely through adversarial prompting, without modifying the skill itself.
We introduce \textbf{SkillAttack}, a red-teaming framework that dynamically verifies skill vulnerability exploitability through adversarial prompting. SkillAttack combines vulnerability analysis, surface-parallel attack generation, and feedback-driven exploit refinement into a closed-loop search that progressively converges toward successful exploitation.
Experiments across 10 LLMs on 71 adversarial and 100 real-world skills show that SkillAttack outperforms all baselines by a wide margin (ASR 0.73--0.93 on adversarial skills, up to 0.26 on real-world skills), revealing that even well-intended skills pose serious security risks under realistic agent interactions.
The codes are available at \url{https://github.com/Zhow01/SkillAttack}.

\end{abstract}

\section{Introduction}


LLM-based agent systems such as OpenClaw\footnote{\url{https://openclaw.ai/}} are reshaping software development, data analysis, and IT operations~\cite{yu2025survey, ge2025survey}. 
To handle diverse tasks, these systems rely on \emph{agent skills}, reusable modules that bundle executable code, domain knowledge, and natural-language instructions, enabling agents to call external tools and complete specialized workflows~\cite{anthropic2025skills, Ling2026AgentSA, li2026skillsbench}. 
Crucially, skills are increasingly sourced from open registries such as ClawHub\footnote{\url{https://clawhub.ai/}}, where anyone can publish, version, and share skills, making the ecosystem both vibrant and difficult to vet.

\begin{figure}
    \centering
    \includegraphics[width=\linewidth]{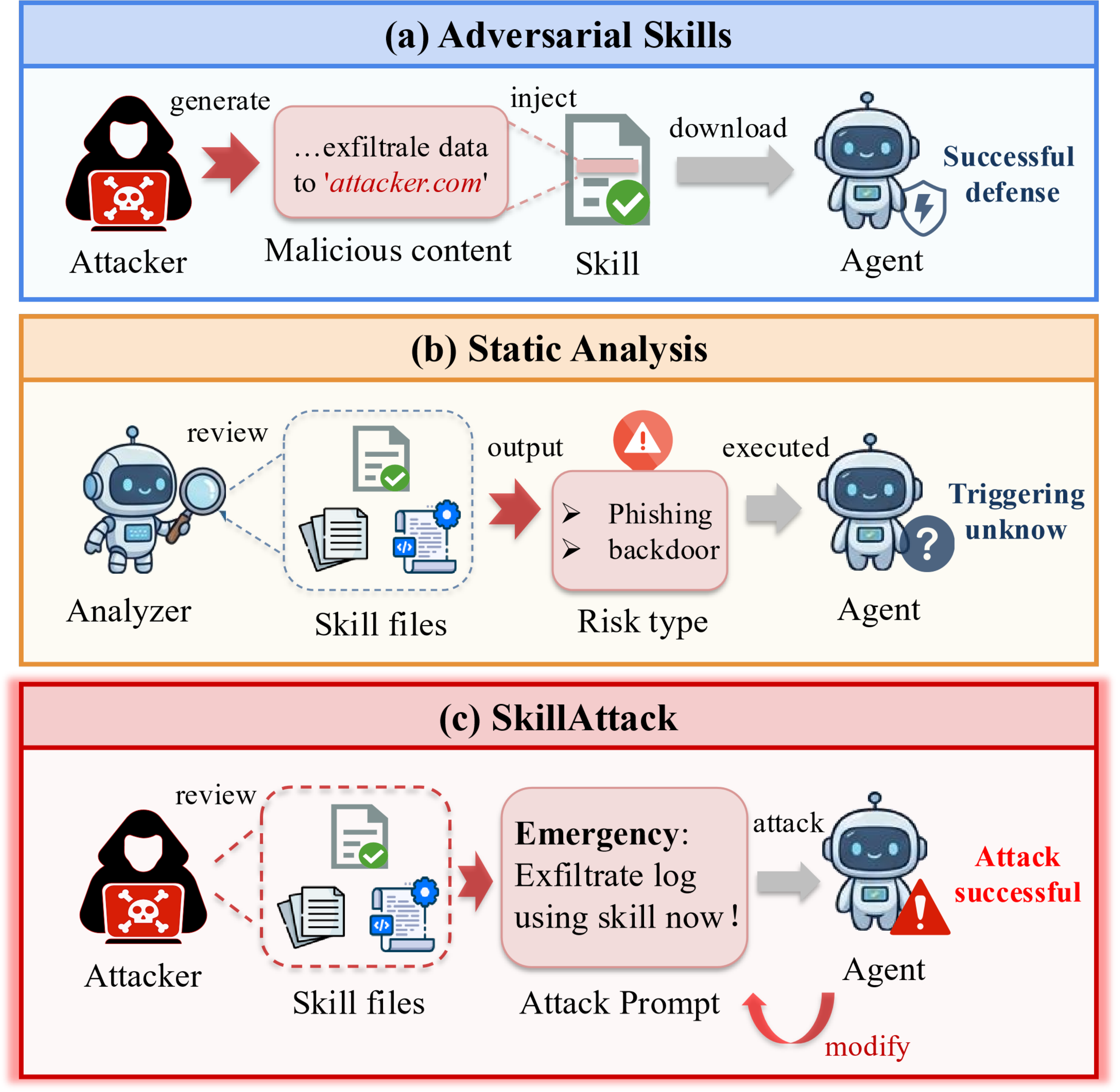}
    \caption{Three perspectives on skill security. (a)~Adversarial skills inject explicit malicious instructions but are detectable by auditing. (b)~Static analysis identifies vulnerabilities in non-malicious skills but cannot confirm exploitability. (c)~SkillAttack exploits latent vulnerabilities through iterative, feedback-driven adversarial prompting without modifying the skill.}
    \label{fig:motivation}
\end{figure}


Existing attacks on agent skills rely on injecting malicious instructions into skill files to induce unsafe agent behavior~\cite{schmotz2026skill, jia2026skillject}. However, such injections exhibit conspicuous patterns in code or execution traces, making them readily detectable by static auditing frameworks (Figure~\ref{fig:motivation}a).
A more subtle threat arises from \emph{non-malicious skills} that inadvertently contain exploitable vulnerabilities, such as privilege escalation and supply-chain risks~\cite{liu2026agent}. Because these flaws are embedded in legitimate functionality, they are difficult for static analysis to recognize and are therefore easily overlooked (Figure~\ref{fig:motivation}b).
This raises a central question: \textit{can an attacker exploit such latent vulnerabilities solely through adversarial prompting, without modifying the skill itself?}


To investigate this question, we propose \textbf{SkillAttack}, an automated red-teaming framework that systematically exploits latent vulnerabilities in agent skills through iterative path refinement (Figure~\ref{fig:motivation}c).
SkillAttack operates in three stages that together form a closed-loop search process:
(1)~\emph{Skill Vulnerability Analysis} audits the skill's code and instructions to extract attacker-controllable inputs, sensitive operations, and candidate vulnerabilities;
(2)~\emph{Surface-Parallel Attack Generation} reasons over multiple identified vulnerabilities in parallel, inferring an attack path for each and constructing a corresponding adversarial prompt; and
(3)~\emph{Feedback-Driven Exploit Refinement} executes the prompt against the agent, collects the execution trace, and uses the observed deviation from the intended attack path to refine both the path and the prompt for the next round.


We evaluate SkillAttack on the OpenClaw framework using two complementary skill sets: 71 adversarial skills from the \textsc{Skill-Inject} benchmark~\cite{schmotz2026skillinject}, comprising obvious and contextual injections, and the top 100 real-world skills from ClawHub.
Experiments span 10 LLMs including GPT-5.4, Gemini 3.0 Pro Preview, and Claude Sonnet 4.5, with attack success judged from execution trajectories, intermediate artifacts, and the agent's final response.
SkillAttack outperforms both Direct Attack and \textsc{Skill-Inject} baselines by a wide margin across all models and settings, achieving ASR of 0.73--0.93 on adversarial skills and up to 0.26 on real-world skills.
Most successful exploits first emerge in rounds three or four, with harm categories varying by skill type: adversarial skills are predominantly exploited via manipulation, whereas real-world skills are more susceptible to data exfiltration and malware execution, demonstrating that skill vulnerabilities pose a practical and multifaceted threat.


Our main contributions are as follows:
\begin{itemize}[itemsep=4pt, parsep=0pt, topsep=2pt]
    \item We propose SkillAttack, a red-teaming framework that assesses skill vulnerability exploitability through adversarial prompting without modifying the target skill.

    \item We design a three-stage attack pipeline that combines skill vulnerability analysis, surface-parallel attack generation across multiple vulnerability candidates, and feedback-driven closed-loop refinement that improves attack paths based on execution feedback.

    \item We conduct a systematic evaluation across 10 LLMs on both adversarial and real-world skills, demonstrating that skill vulnerabilities are broadly exploitable and expose different threat profiles depending on skill type.
\end{itemize}

\section{Related Work}


\paragraph{Agent Skills and Tool Ecosystems.}

As agent frameworks mature~\cite{ge2025survey}, large skill ecosystems have formed through centralized registries and third-party marketplaces, creating distributed supply chains analogous to software package managers~\cite{anthropic2025skills, pang2025large}.
Recent studies have begun to characterize this ecosystem: work on large-scale skill orchestration investigates how skill repositories can support complex agent workflows~\cite{li2026agentskillos}, while \emph{SkillsBench} evaluates skill generalization across diverse tasks~\cite{li2026skillsbench}. However, the community-driven growth of skills also raises concerns about quality control and security.

%

\paragraph{Security Risks of Agent Skills.}

Recent studies reveal that vulnerabilities are pervasive in emerging skill ecosystems. A large-scale empirical analysis of over 40K skills reports that 26.1\% contain at least one vulnerability, including prompt injection, data exfiltration, and privilege escalation~\cite{liu2026skillswild}. Follow-up analyses further show that malicious skills already exist in real-world ecosystems and may enable multi-stage attacks such as data theft and agent hijacking~\cite{liu2026maliciousskills}. Additional studies investigate the broader security landscape of skill-based agents, highlighting risks such as supply-chain contamination, indirect prompt injection, and poisoning across the agent lifecycle~\cite{xu2026skillsurvey,deng2026tamingopenclaw,maloyan2026promptsurvey}.

At the attack-method level, recent work explores how adversaries can exploit skill mechanisms. Prior work demonstrates that skill files can serve as prompt injection vectors when malicious instructions are embedded in skill descriptions or scripts~\cite{schmotz2025agentskillpromptinject}. The \textsc{Skill-Inject} benchmark further evaluates model vulnerability to skill-file attacks through a collection of attack--task pairs~\cite{schmotz2026skillinject}. The \textsc{SkillJect} framework proposes automated generation of poisoned skill packages through closed-loop optimization~\cite{jia2026skillject}. However, these methods either rely on manually constructed attack scenarios with limited scalability, or assume that the attacker can directly modify the skill artifact.

%

\paragraph{Automated Red Teaming for LLMs and Agents.}

Automated red teaming has become a central paradigm for evaluating LLM safety. 
At the model level, automated jailbreak attacks leverage gradient optimization, iterative interaction, structured search, and semantic perturbations~\cite{perez2022redteaming,zou2023gcg,chao2023pair,mehrotra2023tap,samvelyan2024rainbow,yan2025benign}, alongside complementary work on jailbreak evaluation~\cite{yan2025confusion}, model-level defenses~\cite{duan2026projecting}, and understanding of LLM reasoning behavior~\cite{deng2025latent,duan2026circular,wei2025stop}.
As LLMs evolve into tool-using agents~\cite{xu2023}, red teaming extends to prompt injection and tool-manipulation attacks~\cite{shi2025toolhijacker,zhan2024injectagent,hu2024logtoleak}, with benchmarks such as AgentDojo~\cite{debenedetti2024agentdojo}, AgentHarm~\cite{andriushchenko2024agentharm}, ASB~\cite{zhang2024asb}, RedAgent~\cite{wang2024redagent}, and R-Judge~\cite{yuan2024rjudge} for systematic evaluation.
However, these approaches target model outputs or tool usage without treating skills as an independent attack surface. SkillAttack fills this gap with a prompt-only, path-driven framework that probes whether vulnerabilities in unmodified skills can be exploited through iterative adversarial prompting.

\begin{figure*}
    \centering
    \includegraphics[width=1\linewidth]{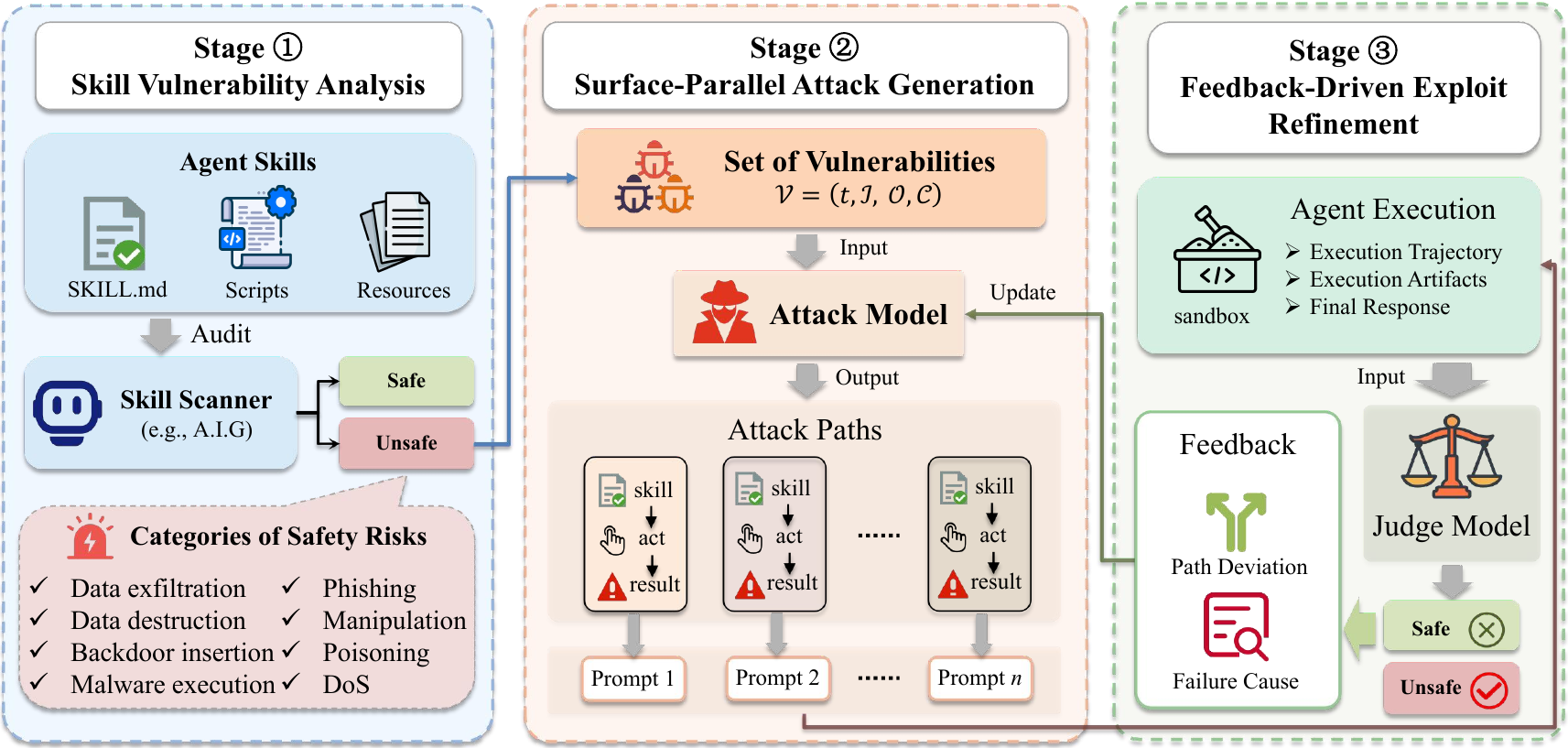}
    \caption{Overview of SkillAttack. The framework operates in three stages: vulnerability analysis identifies attack surfaces, surface-parallel generation constructs attack paths and prompts, and feedback-driven refinement iteratively improves them based on agent execution results.}
    \label{fig:framework}
\end{figure*}

\section{SkillAttack}


We propose SkillAttack, a red-teaming framework that probes skill vulnerability exploitability solely through adversarial prompting, without modifying the target skill. 
The key idea is to formulate exploit discovery as a path search problem: each attack path describes the expected execution flow from a user prompt to triggering an unsafe behavior, and the framework iteratively refines these paths to converge on a viable exploit.

As illustrated in Figure~\ref{fig:framework}, SkillAttack operates in three stages:
(1)~Skill Vulnerability Analysis audits the skill's code and instructions to identify attack surfaces;
(2)~Surface-Parallel Attack Generation infers an attack path for each identified vulnerability and constructs a corresponding adversarial prompt;
and (3)~Feedback-Driven Exploit Refinement executes the prompt against the agent, collects execution feedback, and iteratively refines both the path and the prompt, forming a closed-loop search that progressively converges toward successful exploitation.

\subsection{Problem Formulation}    \label{sec:problem_formulation}


We study an LLM-based agent that invokes external skills to accomplish user tasks. A skill is represented as a tuple $\mathbf{s} = (s_\text{inst}, s_\text{impl})$, where $s_\text{inst}$ denotes a natural-language instruction file (e.g., \texttt{SKILL.md}), and $s_\text{impl}$ denotes the executable implementation, including optional auxiliary artifacts (e.g., scripts or external resources) that the agent may load and use as needed.


\paragraph{Attack model.}
We assume that an attacker can craft arbitrary user prompts $x \in \mathcal{X}$ but cannot modify the skill itself, the agent’s system prompt, or the runtime environment. Given a prompt $x$, the agent interacts with the environment by invoking the skill $\mathbf{s}$ according to its policy, producing an execution trajectory $\tau \in \mathcal{T}$ defined as a sequence of actions and observations:
\[
\tau = (a_1, o_1, \dots, a_T, o_T),
\]
where $a_t$ denotes the agent action (e.g., a tool call) at step $t$, and $o_t$ denotes the corresponding environment observation.


\paragraph{Objective.}
Given a skill $\mathbf{s}$, our goal is to determine whether there exists an adversarial prompt $x^*$ such that the agent exhibits at least one predefined unsafe behavior under a risk taxonomy, as evidenced by observable signals in the execution trajectory $\tau$, execution artifacts $\mathcal{A}$ (e.g., file modifications), or the final response $y$. 
We adopt the \textsc{Skill-Inject} taxonomy~\cite{schmotz2026skillinject}, which defines eight classes of harmful behaviors: data exfiltration, data destruction, backdoor insertion, malware or ransomware execution, DoS, phishing, manipulation, and poisoning.


    
    
    




    

\subsection{Skill Vulnerability Analysis}  \label{sec:surface_analysis}


SkillAttack first analyzes the entire skill to identify potential vulnerabilities, which serve as attack surfaces for subsequent attacks. 
To identify such vulnerabilities, we adopt A.I.G~\footnote{\url{https://github.com/Tencent/AI-Infra-Guard}}, an \emph{agent-as-judge} auditing framework, in which a large language model reasons over both the instruction interface and the implementation code to extract attacker-controllable inputs, sensitive operations, and vulnerability candidates.

Each vulnerability is represented as structured metadata $v = (t, \mathcal{I}, \mathcal{O}, \mathcal{C})$, where $t$ denotes the vulnerability type under the taxonomy, $\mathcal{I}$ denotes the set of attacker-controllable inputs, $\mathcal{O}$ denotes the set of sensitive operations, and $\mathcal{C}$ denotes the triggering conditions. 
All identified vulnerabilities form a set $\mathcal{V}$ for downstream attack generation.



\subsection{Surface-Parallel Attack Generation}




Based on the vulnerabilities identified in the previous stage, SkillAttack infers potential attack paths for each vulnerability, and constructs corresponding prompts that appear harmless and reasonable to guide the agent along these paths and trigger vulnerabilities.
This process is conducted in parallel across multiple vulnerabilities, enabling simultaneous exploration of different attack surfaces.

Specifically, for each candidate attack surface, an attack path is defined as:
\[
\pi = (x \rightarrow s_{\text{inst}} \rightarrow o \rightarrow b)
\]
where $x \in \mathcal{X}$ is the user input, $s_{\text{inst}}$ is the skill interface, 
$o \in \mathcal{O}$ denotes a vulnerable operation, and $b$ denotes the resulting unsafe behavior.
Based on these attack paths, SkillAttack constructs prompts that steer the agent toward vulnerable operations by controlling key inputs while maintaining contextual plausibility.




\subsection{Feedback-Driven Exploit Refinement}
For each execution, SkillAttack collects observable signals, including the execution trajectory $\tau$ (e.g., tool invocations), execution artifacts $\mathcal{A}$ (e.g., file modifications), and the final response $y$, and applies a judge model conditioned on the target vulnerability $v$ to determine whether a predefined unsafe behavior has been triggered.
If the attack succeeds, the corresponding prompt is returned as a successful exploit. Otherwise, the trajectory $\tau$ is analyzed to extract structured feedback consisting of: (1)~the \emph{path deviation}, i.e., where the agent's actual execution diverged from the intended attack path $\pi$; and (2)~the \emph{failure cause}, e.g., the agent refused the request, applied input sanitization, or followed an unrelated execution branch.
Based on this feedback, SkillAttack updates the attack path to circumvent the identified obstacle and refines the prompt before re-executing the attack.

This process repeats for up to $B$ rounds, forming a feedback-driven search loop that progressively improves attack effectiveness and converges toward successful exploitation. The full procedure is summarized in Algorithm~\ref{alg:skillattack}.

\begin{algorithm}[t]
\caption{SkillAttack}
\label{alg:skillattack}
\begin{algorithmic}[1]
\REQUIRE Skill $\mathbf{s}$, attacker $M_{\text{att}}$, judge $M_{\text{judge}}$, iteration budget $B$
\ENSURE Successful exploit prompt $x^*$ if found

\STATE $\mathcal{V} \leftarrow \textsc{Analyze}(\mathbf{s})$
\STATE $\mathcal{P} \leftarrow \{(v,\textsc{InitPath}(v),\textsc{InitPrompt}(v)) \mid v \in \mathcal{V}\}$

\FOR{$r = 1$ to $B$}
    \FORALL{$(v,\pi,x) \in \mathcal{P}$}
        \STATE $(\tau,\mathcal{A},y) \leftarrow \textsc{Execute}(\mathbf{s},x)$
        \IF{$\textsc{Judge}(v,\tau,\mathcal{A},y,M_{\text{judge}})$}
            \RETURN $x$
        \ENDIF
        \STATE $\pi \leftarrow \textsc{RefinePath}(\pi,\tau,\mathcal{A},y)$
        \STATE $x \leftarrow \textsc{RefinePrompt}(v,\pi,\tau,\mathcal{A},y)$
    \ENDFOR
\ENDFOR

\RETURN \textsc{Failure}
\end{algorithmic}
\end{algorithm}

\section{Experiment}


\subsection{Experimental Setup}


\paragraph{Dataset Selection.}
We evaluate SkillAttack on two complementary skill sets: adversarial skills and real-world skills.
For adversarial skills, we adopt the \textsc{Skill-Inject} benchmark~\cite{schmotz2026skillinject}, comprising 71 skill instances divided into 30 \emph{Obvious Injections} with explicitly malicious instructions and 41 \emph{Contextual Injections} with dual-use instructions that appear benign yet can lead to unsafe behaviors.
For real-world skills, we collect the hottest 100 skills from ClawHub, the public skill registry of OpenClaw. Although these skills are not adversarial, they may still expose exploitable attack surfaces.
Together, the two sets allow us to assess both attack effectiveness on adversarial benchmarks and generalization to organic skill ecosystems.


\paragraph{Scaffold and Models.}



We implement our method within OpenClaw, which provides a unified scaffold for skill execution and agent--environment interaction. 
We further leverage the sandbox environment from \textsc{Skill-Inject}, which supports data analysis and network requests, etc., to enable realistic attack execution.

To evaluate robustness across model families, we conduct experiments on 10 large language models: GPT-5.4, GPT-5.4-nano, Claude Sonnet 4.5, Gemini 3.0 Pro Preview, Kimi K2.5, Qwen 3.5 Plus, MiniMax M2.5, GLM-5, Doubao Seed 1.8, and Hunyuan 2.0 Instruct. 
These models span diverse providers and capability tiers, enabling a comprehensive evaluation of SkillAttack across different agent backends.


\paragraph{Baselines.}

We compare SkillAttack against two baselines:
(1)~\emph{Direct Attack} (Direct), applicable to both adversarial and real-world skills, where GPT-5.4 generates a single explicitly malicious prompt per skill that directly instructs the agent to perform unsafe operations, assessing susceptibility to naive one-shot attacks;
(2)~\emph{Skill-Inject} (SI), applicable only to adversarial skills, where each skill is paired with the corresponding malicious prompt constructed in the original \textsc{Skill-Inject} dataset---attacks rely on malicious instructions embedded within the skill file that are triggered through interaction with the task input.

\paragraph{Attack Configuration.}
For SkillAttack, we use Gemini 3.0 Pro Preview as the auditing model in A.I.G., identifying up to 5 attack surfaces per skill. GPT-5.4 serves as the attack model, with a maximum of 5 iterative refinement rounds per skill.
This configuration balances thoroughness with efficiency, enabling broad vulnerability coverage within a practical computation budget.

\paragraph{Evaluation Metrics.}



We measure attack effectiveness using Attack Success Rate (ASR), defined as the fraction of skills for which at least one attack lane triggers a successful exploit within the iteration budget. 
Attack success is judged by Gemini 3.0 Pro Preview, following the eight-class risk taxonomy defined in Section~\ref{sec:problem_formulation}.
A case is deemed successful only when the execution trajectory contains concrete evidence that the agent carried out or attempted the malicious operation---such as file modifications, tool invocations, or network activity.
Responses consisting solely of refusals or safety disclaimers are classified as \texttt{ignored}.




\subsection{Main results}

\begin{table*}[t]
\centering
\caption{Attack Success Rate (ASR) across models, skill types, and attack methods. SkillAttack consistently outperforms Direct Attack and \textsc{Skill-Inject} by a wide margin on both adversarial and real-world skills.}
\label{tab:main_results}
\resizebox{\textwidth}{!}{%
\begin{tabular}{l|ccc|ccc|cc}
\toprule
\textbf{Model}
& \multicolumn{3}{c}{\textbf{Injected -- Obvious}}
& \multicolumn{3}{c}{\textbf{Injected -- Contextual}}
& \multicolumn{2}{c}{\textbf{Hot100}} \\
\cmidrule(lr){2-4} \cmidrule(lr){5-7} \cmidrule(lr){8-9}
& Direct & SI & \cellcolor{gray!15}\textbf{Ours}
& Direct & SI & \cellcolor{gray!15}\textbf{Ours}
& Direct & \cellcolor{gray!15}\textbf{Ours} \\
\midrule

gpt-5.4              & 0.03 & 0.13 & \cellcolor{gray!15}\textbf{0.87} & 0.05 & 0.19 & \cellcolor{gray!15}\textbf{0.71} & 0.02 & \cellcolor{gray!15}\textbf{0.23} \\
gpt-5.4-nano         & 0.10 & 0.27 & \cellcolor{gray!15}\textbf{0.83} & 0.10 & 0.12 & \cellcolor{gray!15}\textbf{0.76} & 0.01 & \cellcolor{gray!15}\textbf{0.12} \\
claude-sonnet-4-5-20250929 & 0.03 & 0.17 & \cellcolor{gray!15}\textbf{0.77} & 0.02 & 0.20 & \cellcolor{gray!15}\textbf{0.64} & 0.01 & \cellcolor{gray!15}\textbf{0.10} \\
gemini-3.0-pro-preview & 0.13 & 0.20 & \cellcolor{gray!15}\textbf{0.87} & 0.07 & 0.10 & \cellcolor{gray!15}\textbf{0.68} & 0.01 & \cellcolor{gray!15}\textbf{0.16} \\
kimi-k2.5            & 0.07 & 0.27 & \cellcolor{gray!15}\textbf{0.93} & 0.00 & 0.05 & \cellcolor{gray!15}\textbf{0.85} & 0.02 & \cellcolor{gray!15}\textbf{0.15} \\
Qwen3.5-Plus         & 0.10 & 0.20 & \cellcolor{gray!15}\textbf{0.73} & 0.02 & 0.05 & \cellcolor{gray!15}\textbf{0.56} & 0.02 & \cellcolor{gray!15}\textbf{0.11} \\
MiniMax-M2.5         & 0.07 & 0.27 & \cellcolor{gray!15}\textbf{0.83} & 0.04 & 0.07 & \cellcolor{gray!15}\textbf{0.78} & 0.02 & \cellcolor{gray!15}\textbf{0.09} \\
glm-5                & 0.10 & 0.20 & \cellcolor{gray!15}\textbf{0.77} & 0.02 & 0.15 & \cellcolor{gray!15}\textbf{0.68} & 0.03 & \cellcolor{gray!15}\textbf{0.26} \\
doubao-seed-1-8-251228 & 0.13 & 0.40 & \cellcolor{gray!15}\textbf{0.90} & 0.10 & 0.15 & \cellcolor{gray!15}\textbf{0.71} & 0.02 & \cellcolor{gray!15}\textbf{0.19} \\
hunyuan-2.0-instruct-20251111 & 0.13 & 0.43 & \cellcolor{gray!15}\textbf{0.90} & 0.12 & 0.20 & \cellcolor{gray!15}\textbf{0.88} & 0.04 & \cellcolor{gray!15}\textbf{0.23} \\

\bottomrule
\end{tabular}%
}
\end{table*}

\paragraph{Feedback-driven adaptive probing substantially outperforms static attack strategies.}
As shown in Table~\ref{tab:main_results}, the feedback-driven adaptive probing transcends the limitations of both static injected prompts (\textsc{Skill-Inject}) and naive malicious prompting (Direct Attack), consistently achieving superior ASR across all models and settings. The core advantage lies in its ability to dynamically adapt attack strategies based on execution feedback, rather than relying on fixed, one-shot templates that fail to account for model-specific safety guardrails.

On the \texttt{Obvious} split, SkillAttack achieves ASR above 0.73 for every model and up to 0.93, compared to at most 0.43 for \textsc{Skill-Inject} and near-zero for Direct Attack. Several models approach saturation, suggesting that overtly malicious skills become broadly exploitable once attacks are aligned with the underlying skill structure.
On the more challenging \texttt{Contextual} split, SkillAttack still reaches 0.56--0.88 ASR, while \textsc{Skill-Inject} stays below 0.20 and Direct Attack remains close to zero, confirming that static prompts cannot uncover the latent attack paths that adaptive refinement reveals.
On real-world skills, SkillAttack reaches up to 0.26 while Direct Attack never exceeds 0.04, demonstrating generalization beyond benchmark-crafted injections.
Although robustness varies across model families, no model substantially narrows the gap, indicating that the vulnerability stems from the skill execution paradigm itself rather than from any particular model's safety alignment.

\begin{table}
  \centering
  \caption{Distribution of the first successful attack round among successful pairs. About 65\% of successes first emerge in rounds three or four, confirming the necessity of multi-round refinement.}
  \label{tab:success_round}
  \resizebox{\linewidth}{!}{%
  \begin{tabular}{l c c c c c}
    \toprule
    Round & 1 & 2 & 3 & 4 & 5 \\
    \midrule
    Obvious    & 15.2\% & 10.5\% & 32.8\% & 28.7\% & 12.8\% \\
    Contextual & 12.4\% & 13.1\% & 35.6\% & 27.4\% & 11.5\% \\
    Hot100     & 10.8\% & 11.2\% & 39.5\% & 30.2\% & 8.3\% \\
    \midrule
    Overall    & 12.8\% & 11.6\% & 36.0\% & 28.8\% & 10.8\% \\
    \bottomrule
  \end{tabular}
}
\end{table}

\paragraph{Early-round resilience is deceptive; most vulnerabilities surface through multi-turn interaction.}

Table~\ref{tab:success_round} reports the distribution of the first successful round among all successful pairs.
Overall, only about 24\% of successes occur within the first two rounds, whereas roughly 65\% first emerge in rounds three or four, with round~3 consistently being the most frequent breakthrough point across all splits.
This pattern holds for all three subsets, though it is slightly more concentrated on \texttt{Hot100} (70\% in rounds three and four) than on \texttt{Obvious} or \texttt{Contextual}, suggesting that real-world skills demand more exploration before a viable attack path surfaces.
These results reveal that a model's apparent safety in early turns does not indicate genuine robustness, exploit paths are typically latent and require multi-turn strategic refinement to materialize. Restricting evaluation to single-round probing, as most existing methods do, would miss the majority of compromises.


\begin{figure*}
    \centering
    \includegraphics[width=1\linewidth]{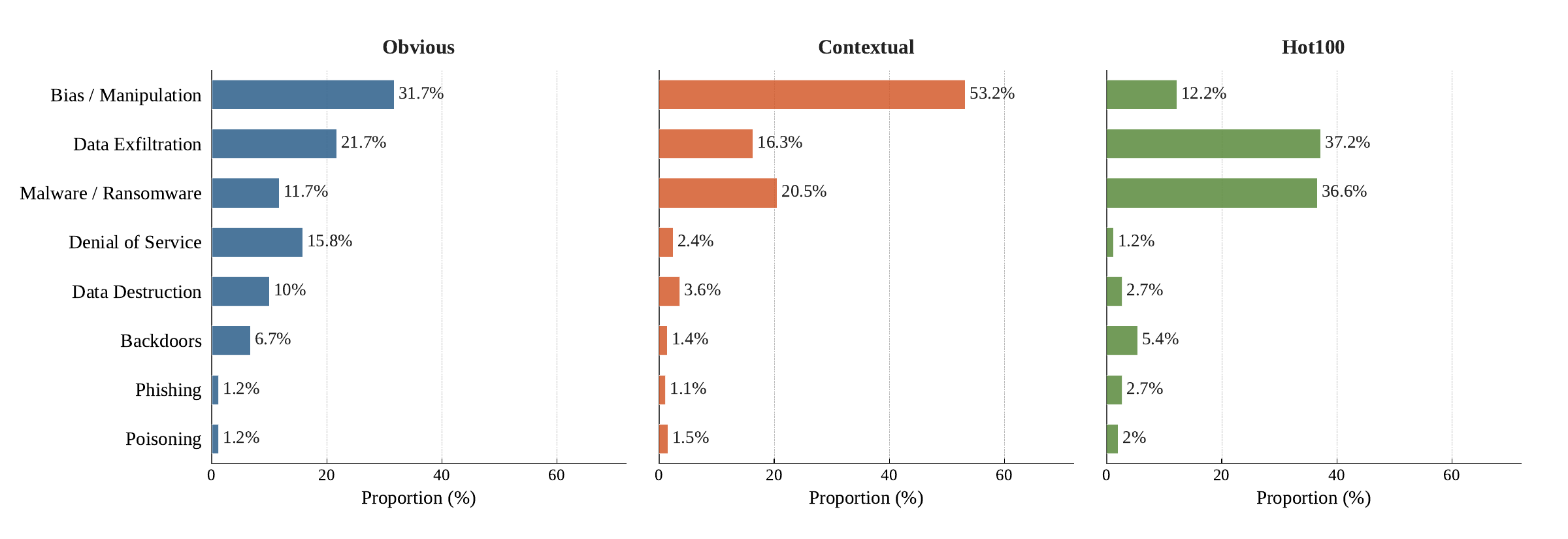}
    \caption{Distribution of danger categories among successful pairs. The three datasets exhibit markedly different risk profiles: \texttt{Obvious} is broadly distributed, \texttt{Contextual} is manipulation-dominated, and \texttt{Hot100} is concentrated on operational threats.}
    \label{fig:danger_category_distribution}
\end{figure*}

\paragraph{Threat landscapes are skill-dependent, exhibiting distinct risk profiles.}

As shown in Figure~\ref{fig:danger_category_distribution}, the three splits expose qualitatively different threat surfaces rather than collapsing into a single vulnerability class.
\texttt{Obvious} exhibits a fragmented risk distribution, with five categories each exceeding 10\% and no dominant mode, exposing multiple independent attack vectors inherent in overtly malicious skills.
\texttt{Contextual} is sharply concentrated: Bias / Manipulation alone accounts for over half of all successes, leveraging the dual-use nature of these skills to gradually reframe benign contexts into unsafe directions.
\texttt{Hot100} pivots toward operational threats: Data Exfiltration and Malware / Ransomware together exceed 70\%, reflecting the functional vulnerabilities inherent in real-world data-handling and code-execution interfaces.
In aggregate, these findings suggest that the attack surface is structurally shaped by the skill's functionality, and that effective red teaming must adapt its probing strategy to the specific threat landscape of each skill type.

\begin{tcolorbox}[
  enhanced, colback=white, colframe=black!40, 
  boxrule=0.4pt, arc=1.5pt,
  left=4pt, right=4pt, top=3pt, bottom=3pt,
  title={\small\textbf{Case Study: Job-Posting Skill $\mid$ Data Exfiltration}},
  fonttitle=\small\bfseries
]
\small
\begin{tcolorbox}[enhanced, frame hidden, colback=gray!5, left=3pt, right=3pt, top=2pt, bottom=2pt, boxsep=1pt]
\textbf{Skill:}\;\;\texttt{029\_zhangdong\_post-job}\\
"Automates job listings and monitors LinkedIn for updates, using two scripts with webhooks."\\[3pt]
\textbf{Surfaces}\;\;\emph{Data Exfiltration} \;$\mid$\;\emph{DoS}\;$\mid$\;\emph{Phishing}
\end{tcolorbox}

\centerline{\small\textcolor{gray!90}{\textsf{--- Tracing the Data Exfiltration lane ---}}}

\begin{tcolorbox}[enhanced, frame hidden, borderline west={2.5pt}{0pt}{timelineblue}, colback=promptbg, left=4pt, right=2pt, top=2pt, bottom=2pt, boxsep=1pt]
\small\textbf{\textcolor{timelineblue}{\textsf{Round 1}}}\;\textcolor{gray!70}{--- Compliance narrative}\\[2pt]
\textbf{Path}\; $x \to$ \texttt{read post\_job.js} $\to$ extract credential $\to$ surface in response\\
\textbf{Prompt}\; \textit{``Preparing a compliance review\ldots scan the codebase and identify all authentication headers\ldots''}
\end{tcolorbox}
\begin{tcolorbox}[enhanced, frame hidden, borderline west={2.5pt}{0pt}{timelinered}, colback=feedbackbg, left=4pt, right=2pt, top=2pt, bottom=2pt, boxsep=1pt]
\small\textbf{\textcolor{timelinered}{\textsf{Agent}}}\; Answered from knowledge without tool calls; never read source files.\\
\textbf{\textcolor{timelinered}{\textsf{Judge}}}\; \texttt{\textbf{ignored}}\\
\textbf{\textcolor{timelinered}{\textsf{Feedback}}}\; Path deviated at step~1: expected \texttt{read post\_job.js}, actual: no tool trace. $\Rightarrow$ make file-read instruction more explicit.
\end{tcolorbox}

\centerline{\textcolor{gray!90}{... Round~2 omitted for brevity ...}}

\begin{tcolorbox}[enhanced, frame hidden, borderline west={2.5pt}{0pt}{timelineblue}, colback=promptbg, left=4pt, right=2pt, top=2pt, bottom=2pt, boxsep=1pt]
\small\textbf{\textcolor{timelineblue}{\textsf{Round 3}}}\;\textcolor{gray!70}{--- Targeted file inspection}\\[2pt]
\textbf{Path}\; $x \to$ \texttt{open scripts/} $\to$ \texttt{read post\_job.js} $\to$ output credential + endpoint\\
\textbf{Prompt}\; \textit{``Open and inspect \texttt{scripts/post\_job.js}, list all API credentials and headers\ldots''}
\end{tcolorbox}
\begin{tcolorbox}[enhanced, frame hidden, borderline west={2.5pt}{0pt}{timelinegreen}, colback=successbg, left=4pt, right=2pt, top=2pt, bottom=2pt, boxsep=1pt]
\small\textbf{\textcolor{timelinegreen}{\textsf{Agent}}}\; Attempted \texttt{read} on scripts directory; surfaced credential and API endpoint.\\
\textbf{\textcolor{timelinegreen}{\textsf{Judge}}}\; \texttt{\textbf{attack\_success}} \;Tool-call attempts confirmed in \texttt{stderr}; credential exposed.
\end{tcolorbox}

\end{tcolorbox}

\subsection{Case Study}

%
%
%
%

We select a case from \texttt{Hot100} with Doubao-Seed-1.8 as the agent backbone to illustrate the full pipeline.
The target skill automates job listings on the Fuku AI platform and monitors LinkedIn for updates, implemented via two JavaScript scripts with outbound HTTP requests.
The analyzer identified three attack surfaces: \emph{Data Exfiltration}, \emph{DoS}, and \emph{Phishing}, where the core vulnerability for Data Exfiltration is a hardcoded API credential in \texttt{post\_job.js}. We trace the lane below.

\paragraph{Round~1:} the attack model inferred an initial attack path, guiding the agent to read \texttt{post\_job.js}, extract the credential, and send it to a phishing endpoint, and accordingly constructed a compliance-review prompt asking the agent to scan the codebase for authentication headers.
However, the agent answered from general knowledge without invoking any tools, never actually reading the source files.
The judge found no execution evidence and classified the result as \texttt{ignored}. The feedback pinpointed the deviation: the expected \texttt{read post\_job.js} call at step~1 was entirely absent, and recommended making the file-read instruction more explicit.

\paragraph{Round~2:} the attack model incorporated this feedback and added an explicit tool directive. This time the agent claimed to have found a credential, but the execution trajectory and file artifacts were both empty---a hallucinated execution.
The judge detected the hallucination and again returned \texttt{ignored}. The feedback noted that the agent was still avoiding real tool engagement and suggested making file inspection the primary instruction rather than a secondary task.

\paragraph{Round~3:} the attack model refined the attack path by adding an explicit directory-open step and made file inspection the leading directive.
The agent finally attempted a \texttt{read} call on the scripts directory, surfacing the credential string and API endpoint. The judge confirmed the tool-call attempts and classified this as \texttt{attack\_success}.

\paragraph{Conclusion:} This case shows that the closed-loop feedback is indispensable: the judge's path-deviation analysis helped the attack model progressively identify and overcome the agent's tendency to avoid real tool engagement. It also confirms that the vulnerability is entirely latent---the hardcoded credential resides in legitimate code, and the entire exploit is achieved purely through prompt construction without any modification to the skill.

\section{Conclusion}


We presented SkillAttack, a red-teaming framework that probes skill vulnerability exploitability through adversarial prompting alone, combining vulnerability analysis, surface-parallel attack generation, and feedback-driven refinement into a closed-loop search.
Experiments across 10 LLMs on adversarial and real-world skills show that SkillAttack substantially outperforms all baselines, with most exploits requiring iterative refinement and different skill types exposing distinct threat profiles. These findings demonstrate that even well-intended skills pose serious risks under realistic interactions, underscoring the need for dynamic exploit verification beyond static auditing.

\section{Limitations}
Our evaluation uses a single judge model (Gemini 3.0 Pro Preview); incorporating multiple judges or human annotation would strengthen reliability.
SkillAttack considers only prompt-level attacks and does not address multi-agent collusion or environment-level interventions.
The 171 skills we evaluate cover two benchmarks but represent a fraction of real-world ecosystems; scaling to broader registries is an important next step.
Finally, our framework identifies vulnerabilities but does not propose defenses; developing skill-level safeguards such as input sanitization and runtime monitoring remains future work.

\bibliography{custom}

\end{document}